# DeepProteomics: Protein family classification using Shallow and Deep Networks

**Anu Vazhayil, Vinayakumar R and Soman KP**
**Center for Computational Engineering and Networking (CEN),Amrita School of Engineering, Coimbatore, Amrita Vishwa Vidyapeetham, India**
**Email: anuv1994@gmail.com, vinayakumarr77@gmail.com, kp soman@amrita.edu**

## ABSTRACT

The knowledge regarding the function of proteins is necessary as it gives a clear picture of biological processes. Nevertheless, there are many protein sequences found and added to the databases but lacks functional annotation. The laboratory experiments take a considerable amount of time for annotation of the sequences. This arises the need to use computational techniques to classify proteins based on their functions. In our work, we have collected the data from Swiss-Prot containing 40433 proteins which is grouped into 30 families. We pass it to recurrent neural network(RNN), long short term memory(LSTM) and gated recurrent unit(GRU) model and compare it by applying trigram with deep neural network and shallow neural network on the same dataset. Through this approach, we could achieve maximum of around 78% accuracy for the classification of protein families.

## 1 INTRODUCTION

Proteins are considered to be essentials of life because it performs a variety of functions to sustain life. It performs DNA replication, transportation of molecules from one cell to another cell, accelerates metabolic reactions and several other important functions carried out within an organism. Proteins carry out these functions as specified by the informations encoded in the genes. Proteins are classified into three classes based on their tertiary structure as globular, membrane and fibrous proteins. Many of the globular proteins are soluble enzymes. Membrane proteins enables the transportation of electrically charged molecules past the cell membranes by providing channels. Fibrous proteins are always structural. Collagen which is a fibrous protein forms the major component of connective tissues. Escherichia coli cell is partially filled by proteins and 3% and 20% fraction of DNA and RNA respectively contains proteins. All of this contributes in making proteomics as a very important field in modern computational biology. It is therefore becoming important to predict protein family classification and study their functionalities to better understand the theory behind life cycle.

Proteins are polymeric macromolecules consisting of amino acid residue chains joined by peptide bonds. And proteome of a particular cell type is a set of proteins that come under the same cell type. Proteins is framed using a primary structure represented as a sequence of 20-letter alphabets which is associated with a particular amino acid base subunit of proteins. Proteins differ from one another by the arrangement of amino acids intent on nucleotide sequence of their genes. This results in the formation of specific 3D structures by protein folding which determines the unique functionality of the proteins. The primary structure of proteins is an abstracted version of the complex 3D structure but retains sufficient information for protein family classification and infer the functionality of the families.

Protein family consists of a set of proteins that exhibits similar structure at sequence as well as molecular level involving same functions. The lack of knowledge of functional information about sequences in spite of the large number of sequences known, led to many works identifying family of proteins based on primary sequences [1-3]. Dayhoff identified the families of numerous proteins [4]. Members of the same protein family can be identified using sequence homology which is defined as the evolutionary relatedness. It also exhibits similar secondary structure through modular protein domains which further group proteins families into super families[5]. These classifications are listed in database like SCOP[6]. Protein family database (Pfam)[7] is an extremely large source which classify proteins into family, domain, repeat or motif. Protein classification using 3D structure is burdensome and require complex techniques like X-ray crystallography and NMR spectroscopy. This led to the works [8-10] which uses only primary structure for protein family classification. In this work we use data from Swiss-Prot for protein family classification and obtain a classification accuracy of about 96%.

In our work we gathered family information of about 40433 protein sequences in Swiss-Prot from Protein family database(Pfam), which consists of 30 distinct families. The application of keras embedding and n-gram technique is used with deep learning architectures and traditional machine learning classifiers respectively for text classification problems in the cyber security[33],[34],[35],[36],[37]. By following, we apply keras word embedding and pass it to various deep neural network models like  recurrent neural network(RNN), long short term memory(LSTM) and gated recurrent unit(GRU) and then compare it performance by applying trigram with deep and shallow neural networks for protein family classification. To verify the model used in our work, we test it over dataset consisting of about 12000 sequences from the same database.

The rest of the part of this paper are organized as follows. Section 2 discusses the related work, Section 3 provides background details of deep learning architecture, Section 4 discusses the proposed methodology, Section 5 provides results and submissions and at last the conclusion and future work directions are placed in Section 6.

## 2 RELATED WORK

There have been many works till date to identify protein functions based on the primary structures aka protein sequences. In this section we describe briefly about the works done in that area.

Needleman[11] along with Wunsch developed an algorithm using dynamic programming which uses global alignment to find similarity between protein and DNA sequences. This method is used when the sequences does not share similar patterns. Whereas in Smith work[12] they used local alignment of protein and DNA sequences and does clustering of protein sequences based on the length of the different fragments in the sequence.

In the current decade people mostly rely on computational techniques like machine learning, deep learning and pattern recognition for the classification of protein families instead of depending on the old techniques which make use of alignment of the sequences. Some of the works which uses machine learning techniques are explained briefly below.

In the works [13-15] primary structure aka protein sequences is used to classify protein families using classifiers like support vector machines(SVM). But apart from protein sequences these methods require additional information for feature extraction. Some of theese are polarity, hydrophobicity, surface tension, normalized Van der Waals volume, charge, polarizability, solvent accessibility and seconday structure which requires a lot of computational power to analyze. In the work [13] protein classification for 54 families achieved 69.1-99.6% accuracy. In another study, Jeong et al. [16] used position-specific scoring matrix(PSSM) for extracting feature from a protein sequence. They used classifiers such as Naive Bayesian(NB), Decision Tree(DT), Support Vector Machine(SVM) and Random Forest(RF) to verify their approach and achieved maximum accuracy of about 72.5%.

Later on hashing was introduced for mapping high dimentional features to low dimentional features using has keys. Caragea et al [17] used this technique to map high dimenstional features obtained through k-gram representation, by storing frequency count of each k-gram in the feature vectors obtained and hashing it together with the same hash key. This method gave accuracy of about 82.83%. Yu et al. [18] proposed a method to represent protein sequences in the form of a k-string dictionary. For this singular value decomposition(SVD) was applied to factorize the probability matrix.

Mikolov et.al. [19] proposed a model architecture to represent words as continous vectors. This approach aka word2vec map words from the lexicon to vectors of real numbers in low dimentional space. When it is trained over a large set of data the linguistic context could be studied from the given data and it will be mapped close to each other in the euclidean space. In [20] they have applied word2vec architecture to the biological sequences. And introduced a new representation called bio-vectors (BioVec) for the biological sequences with ProtVec for protein sequences and GeneVec for gene sequences. The k-mers derived from the data is then given as input to the embedding layer. They achieved family classification accuracy of about 93% by using ProtVec as a dense representation for biologial sequences. In our work, the proposed architecture is trained solely on primary sequence information, achieving a high accuracy when used for classification of protein families.

# 3 BACKGROUND

Text representation aka text encoding can be done in several ways which are mainly of two types, sequential representation and non-sequential representation. Transforming the raw text data to these representations involves preprocessing and tokenizing the texts. During preprocessing, all uppercase characters are changed to lowercase and a dictionary is maintained which assigns a unique identification key to all the characters present in the text corpus. Later we use it in order to map input texts to vector sequence representation. After mapping character with a unique id, vocabulary is created using the training data. The preprocessing of the text data is completed by finally converting all varying length sequences to fixed length sequences.

In this work we represent the text data as a sequence and therefore maintains the word order which incorporates more information to the representation. A network can be modeled by training the data using one’s own representation or by using the existing ones. For our work, we have used keras embedding for text representation. This maps the discrete character ids to its vectors of continuous numbers. The character embedding captures the semantic meaning of the given protein sequence by mapping them into a high dimensional geometric space. This high dimensional geometric space is called as character embedding space. The newly formed continuous vectors are fed to other layers in the network.

Features for text data can be obtained using several techniques, one of them being n-grams. N-grams can be used to represent text data which gives unique meaning when combined together, these combinations are obtained by taking continuous sequences of n characters from the given input sequence. The general equation for the N-gram approximation to the conditional probability of the next word in a sequence is,

$$P(w_n \mid w_1^{n-1}) \approx P(w_n \mid w_{n-N+1}^{n-1})$$

In this work trigrams is used as feature for some of the models. Trigrams is a combination of three adjacent elements of a set of tokens. While computing trigram probability, we use two pseudo-words in the beginning of each sentence to create the first trigram (i.e., P(I | <s><s>).

After the data is preprocessed and represented in the form of continuous vectors, it is fed into other layers like (1) RNN (2) LSTM (3) GRU.

**3.1 Recurrent Neural Network**

RNN was developed to improve the performance of feed forward network(FFN) introduced in 1990 [21]. Both these networks differ by the way they pass the information to the nodes of the network where a series of mathematical operations are performed. FFN pass the information never touching a node twice whereas RNN pass it through a loop and ingesting their own outputs at a later moment as input, hence called recurrent. In a sequence there will be some information and RNN use it to perform the tasks that FNNs fail to do. RNN handles sequence data efficiently for natural language processing(NLP) tasks because it acts on arbitrary length sequence and the unfolded RNN model shares the weight across time steps.

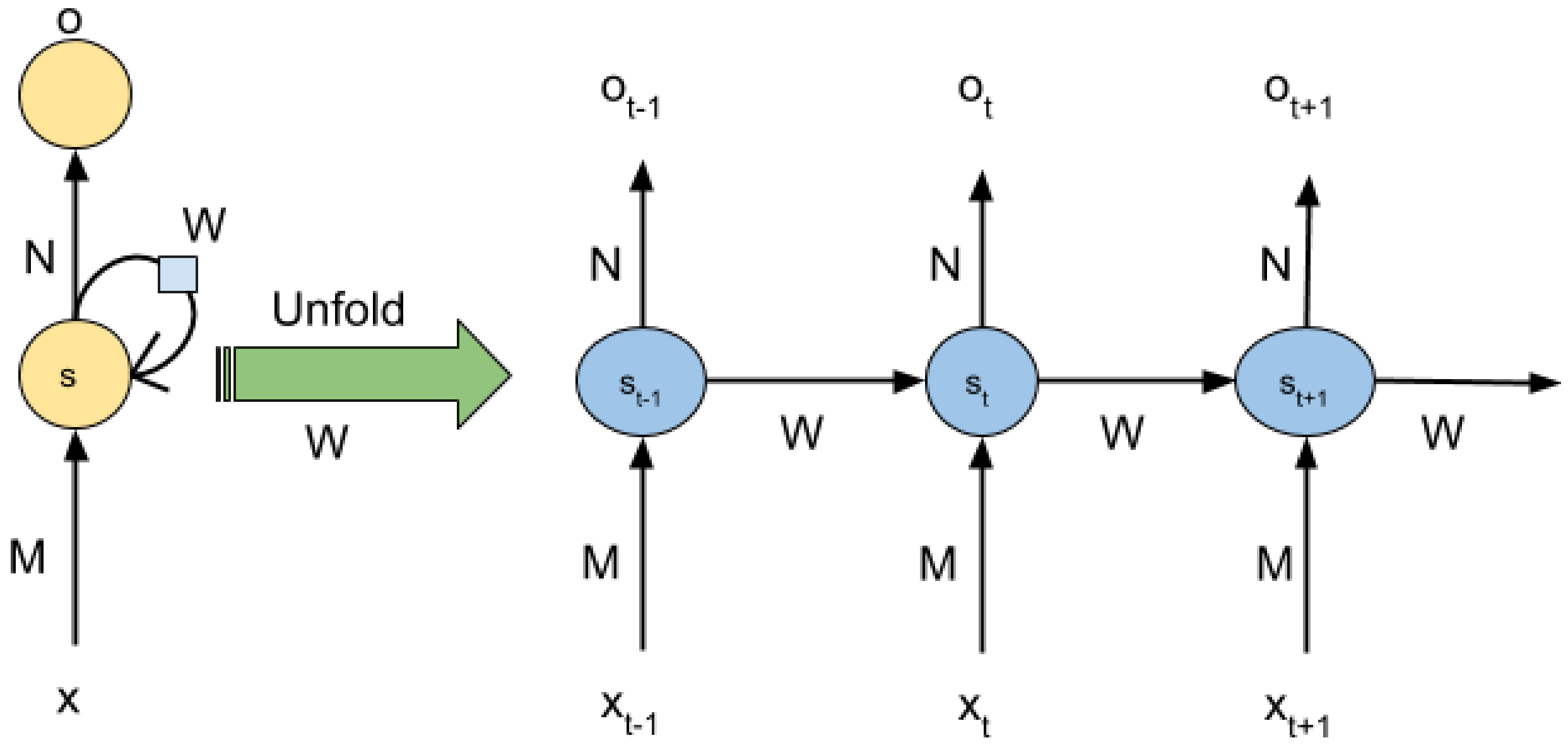

**Fig. 1** Unfolding or unrolling RNN

We can represent mathematically the process of carrying memory forward in a RNN as follows:

$$S_t = f(Ux_t + Ws_{t-1})$$

In the equation, $x_t$ is the input state at time step t and $s_t$ is the hidden state(or memory) at time step t. The function f is a nonlinearity activation function such as tanh or ReLU. During training of long sequences, vanishing and exploding gradient problem will arise due to this form of transition function [22, 23]. To cope up with this issue long short-term memory(LSTM) was introduced[24] using a special unit called memory block. Afterwards many variants to the LSTM architecture was introduced, prominent ones being inclusion of forget gate[25] and peephole connections[26].

RNN shares same parameters(M, N, W in **Fig. 1**) at each layer unlike other traditional deep neural networks. This method reduces the total number of parameters to be learnt. To minimize the loss function an optimal weight parameter (M, N, W) is to be found, using stochastic gradient descent(SGD). Gradient estimation in RNN is done using backpropogation through time(BPTT) [27].

LSTM is an upgraded version of vanilla RNN [28]. Both LSTM and RNN uses backpropagation through time for training the network. While training a traditional RNN, there arises a case where the gradient becomes very small and further learning becomes extremely slow. This happens because the gradient vector can end up multiplied by the weight matrix a large number of times. If the values of the weight matrix is small, then it can lead to vanishing gradient. Similarly if the value of the weight matrix is high, then it can lead to exploding gradient. These problems makes the learning very difficult. The weight matrix plays a major role in training a RNN. These limitations of RNN are the key motivation of LSTM model. The LSTM model introduced the concept of memory cell. The memory cell consists of: an input gate, a neuron with a self recurrent connection, a forget gate and an output gate. In the self recurrence connection of the LSTM network, identity function is used as the activation function and has derivative 1.0. This ensures that the gradient neither explodes nor vanishes since the back-propagated gradient remains constant. Therefore the LSTM is able to learn long term dependencies [29].

Gated Recurrent Units (GRU) is a variant of LSTM recurrent neural networks [30]. Unlike other deep neural networks, GRU and LSTM have parameters specifically to control memory updation. GRU and LSTM are widely used in sequence modelling. They both can capture short term and long term dependencies in sequences. Even though both can be used in sequence modelling, the parameters in a GRU network is less compared to an LSTM network and hence the training in a GRU network is faster when compared to LSTM network. The mathematical expression for a GRU network is as follows:

$$f_t = \boldsymbol{\sigma}_g(W_f x_t + U_f h_{t-1} + b_r)$$
$$h_t = f_t \odot h_{t-1} + (1 - f_t) \odot \boldsymbol{\sigma}_h(W_h x_t + U_h (f_t \odot h_{t-1}) + b_h)$$

In the above equations, $x_t$, $h_t$, $f_t$ represents the input, output, forget vector respectively. And W, U, b are parameter matrices and b bias respectively. Theoretically the reset and forget gate in a GRU network ensures that the memory doesn't get used up by tracing short-term dependencies. In a GRU network the memory is protected by learning how to use its gates, so as to make long term predictions.

# 4 METHODOLOGY

In this section, the performance of deep and shallow neural networks using bigram and RNN, LSTM and GRU with word embedding are evaluated on a data set of protein sequences. First the description of proteins and data set is discussed followed by the proposed architecture.

## 4.1 Protein

Proteins is framed using a primary structure represented as a sequence of 20-letter alphabets which is associated with a particular amino acid base subunit of proteins. Proteins differ from one another by the arrangement of amino acids intent on nucleotide sequence of their genes which gives them different functionalities. Depending on their functions proteins are grouped under different families. Thus able to identify the family of a protein sequence will give us the information of its functions. In our work we have classified protein sequences into 30 families only using the primary structure information.

## 4.2 Dataset

We gathered family information of about 40433 protein sequences in Swiss-Prot from Protein family database(Pfam), which consists of 30 distinct families. Swiss-Prot is a curated database of primary protein sequences which is manually annotated and reviewed. There is no redundancy of protein sequences in the database and is evaluated based on results obtained through experiments. We have divided the obtained protein sequences into 12000 sequences for test data and the rest 28433 protein sequences for training the model. The details of the 30 family names can be found in **Table 1**.

| **Family name** | **Training instances** | **Testing instances** |
|---|---|---|
| 7 transmembrane receptor (rhodopsin family) | 1253 | 567 |
| ATPase family associated with various cellular activities (AAA) | 1291 | 420 |
| Amino acid kinase family | 1370 | 380 |
| ATP synthase alpha-beta family, nucleotide-binding domain | 2057 | 330 |
| Aldehyde dehydrogenase family | 921 | 279 |
| BRCA1 C Terminus (BRCT) domain | 597 | 359 |
| TCP-1-cpn60 chaperonin family | 824 | 422 |
| EPSP synthase (3-phosphoshikimate 1-carboxyvinyltransferase) | 802 | 405 |
| GHMP kinases C terminal domain | 616 | 395 |
| GHMP kinases N terminal domain | 699 | 421 |
| Hsp70 protein | 880 | 392 |
| Helicase conserved C-terminal domain | 2005 | 513 |
| Histidine biosynthesis protein | 846 | 402 |
| KOW motif | 540 | 507 |
| Major Facilitator Superfamily | 915 | 388 |
| MMR_HSR1 | 2606 | 478 |
| Oxidored_q1 | 827 | 424 |
| RF-1 domain | 597 | 353 |
| Ribosomal protein S12-S23 | 668 | 348 |
| Ribosomal protein L16 | 713 | 340 |
| Ribosomal protein L33 | 561 | 397 |
| Ribosomal protein S11 | 577 | 403 |
| Ribosomal protein S14 | 652 | 345 |
| Ribosomal protein S2 | 699 | 384 |
| Ribosomal protein S4 | 691 | 381 |
| Shikimate-quinate 5-dehydrogenase | 705 | 423 |
| Uncharacterized protein family UPF0004 | 625 | 419 |
| UvrB-uvrC motif | 616 | 352 |
| tRNA synthetases class I (I, L, M and V) | 1233 | 401 |
| tRNA synthetases class II (D, K and N) | 1047 | 372 |

**Table 1** Description of data set

### 4.3 Proposed architecture

The proposed architecture typically called as DeepProteomics which composed of Character Embedding, Feature representation, Regularization and Classification sections. Each section is discussed in detail below.

**Character Embedding**

By using the aforementioned approach, a matrix is constructed for training(28433*3988) and testing(12000*3988) for the given dataset. These matrices are then passed to an embedding layer with batch size 128. An embedding layer maps each character onto a 128 length real valued vector. This can be considered as one hyper parameter, we choose 128 to provide further level of freedom to the deep learning architectures. This collaboratively works with other layers in the deep network during backpropagation. This facilitates sequence character clustering and similar characters cluster together. This kind of character clustering facilitates other layers to easily detect the semantics and contextual similarity structures of protein sequences. For comparative study, trigram representation is constructed for protein sequence and using feature hashing approach, the protein sequence lengths are set to 1000.

**Feature representation**

We adopt deep layer RNN for feature representation. Recurrent layer extract sequential information of the protein sequences.

**Recurrent Structures**

We have used RNN, LSTM and GRU as the recurrent structures. In all the experiments, 1 layer of any of the algorithms like RNN, LSTM and GRU is used. The number of units used is 128. The recurrent structuress is followed by a dropout of 0.2 while training. This in turn is followed by fully connected layer with 30 neurons in the output layer.

**Regularization**

A Dropout layer with 0.2 is used in each models between recurrent structures and fully connected layer that acts as a regularization parameter to prevent from overfitting. A Dropout is a method for removing the neurons randomly along with their connections during training a deep learning model.

**Classification**

The embedded character level vectors coordinately works with the recurrent structures to obtain optimal feature representation. This kind of feature representation learns the similarity among the sequences. Finally, the feature representations of recurrent structures is passed to the fully-connected network to compute the probability that the sequence belongs to a particular family. The non-linear activation function in the fully connected layer facilitates in classifying the feature vectors to the respective families. The 1000 length protein sequence vectors are passed as input to shallow DNN, deep DNN and other traditional machine learning classifiers for comparative study.

In fully-connected layer, each neuron in the previous layer has connection to every other neuron in the next layer. It has two layers, a fully connected layer with 128 units followed by fully connected layer with 30 units. In categorizing the proteins to 30 families, the prediction loss of deep learning models is computed using categorical-cross entropy,

$$loss(p,e) = -\sum_x p(x)\log(e(x))$$

Where p is true probability distribution and q is predicted probability distribution. To minimize the loss of categorical-cross entropy we used Adam optimization algorithm[31]. The detailed architecture details of GRU, LSTM and RNN are placed in **Table 2**, **Table 3** and **Table 4** respectively. The detailed architecture of shallow and deep DNN module is given in **Table 5** and **Table 6** respectively.

| **Layers** | **Type** | **Output shape** | **Other parameters** | **Parameters (105,630)** |
|---|---|---|---|---|
| 0-1 | Embedding | (None, 3988, 128) | embedding vector length = 128 | 3072 |
| 1-2 | GRU | (None, 128) | memory blocks 128 | 98688 |
| 2-3 | Dropout | (None, 128) | | 0 |
| 3-4 | Dense | (None, 30) | | 3870 |
| 4-5 | Activation | (None, 30) | softmax | 0 |

**Table 2** Configuration details of proposed GRU Architecture

| **Layers** | **Type** | **Output shape** | **Other parameters** | **Parameters (138,526)** |
|---|---|---|---|---|
| 0-1 | Embedding | (None, 3988, 128) | embedding vector length = 128 | 3072 |
| 1-2 | LSTM | (None, 128) | memory blocks 128 | 131584 |
| 2-3 | Dropout | (None, 128) | | 0 |
| 3-4 | Dense | (None, 30) | | 3870 |
| 4-5 | Activation | (None, 30) | softmax | 0 |

**Table 3** Configuration details of proposed LSTM Architecture

| Layers | Type | Output shape | Other parameters | Parameters (105,630) |
|---|---|---|---|---|
| 0-1 | Embedding | (None, 3988, 128) | embedding vector length = 128 | 3072 |
| 1-2 | GRU | (None, 128) | memory blocks 128 | 98688 |
| 2-3 | Dropout | (None, 128) | | 0 |
| 3-4 | Dense | (None, 30) | | 3870 |
| 4-5 | Activation | (None, 30) | softmax | 0 |

**Table 4** Configuration details of proposed RNN Architecture

| Layers | Type | Output shape | Other parameters | Parameters (105,630) |
|---|---|---|---|---|
| 0-1 | dense_1 (Dense) | (None, 30) | Trigram of length 1000 | 30,030 |

**Table 5** Configuration details of proposed DNN Architecture

| Layers | Type | Output shape | Other parameters | Parameters (144,766) |
|---|---|---|---|---|
| 0-1 | dense_1 (Dense) | (None, 128) | Trigram of length 1000 | 128128 |
| 1-2 | batch_normalization_1 | (None, 128) | | 512 |
| 2-3 | activation_1 (Activation) | (None, 128) | | 0 |
| 3-4 | dropout_1 (Dropout) | (None, 128) | | 0 |
| 4-5 | dense_2 (Dense) | (None, 64) | 64 units | 8256 |
| 5-6 | batch_normalization_2 | (None, 64) | | 256 |
| 6-7 | activation_2 (Activation) | (None, 64) | | 0 |
| 7-8 | dropout_2 (Dropout) | (None, 64) | | 0 |
| 8-9 | dense_3 (Dense) | (None, 64) | 64 units | 4160 |
| 9-10 | batch_normalization_3 | (None, 64) | | 256 |
| 10-11 | activation_3 (Activation) | (None, 64) | | 0 |
| 11-12 | dropout_3 (Dropout) | (None, 64) | | 0 |
| 12-13 | dense_4 (Dense) | (None, 32) | 32 units | 2080 |
| 13-14 | batch_normalization_4 | (None, 32) | | 128 |
| 14-15 | activation_4 (Activation) | (None, 32) | | 0 |
| 15-16 | dropout_4 (Dropout) | (None, 32) | | 0 |
| 16-17 | dense_5 (Dense) | (None, 30) | 30 units | 990 |
| 17-18 | batch_normalization_5 | (None, 30) | | 0 |
| | activation_5 (Activation) | | softmax | |

**Table 6** Configuration details of proposed DNN Architecture

# 5 RESULTS AND ANALYSIS

All the experiments are run on GPU enabled TensorFlow[38] and Keras[39] higher level API. The detailed statistical measures for the protein sequence dataset for the various algorithms used is reported in **Table 5**. The overall performance of the neural network models are better than the traditional machine learning techniques. Thus, we claim that the character level URL embedding with deep learning layers can be a powerful method for automatic feature extraction in the case of protein family classification.

| Algorithm | Accuracy | Precision | Recall | F-score |
|---|---|---|---|---|
| **RNN** | 0.731 | 0.733 | 0.731 | 0.728 |
| **LSTM** | 0.784 | 0.799 | 0.784 | 0.787 |
| **GRU** | 0.772 | 0.787 | 0.772 | 0.774 |
| **Deep DNN** | 0.693 | 0.694 | 0.693 | 0.688 |
| **Shallow DNN** | 0.541 | 0.552 | 0.541 | 0.519 |
| **Logistic regression** | 0.424 | 0.430 | 0.424 | 0.423 |
| **Naive Bayes** | 0.379 | 0.399 | 0.379 | 0.376 |
| **KNN** | 0.262 | 0.310 | 0.262 | 0.263 |
| **Decision tree** | 0.211 | 0.222 | 0.211 | 0.210 |
| **Adaboost** | 0.096 | 0.123 | 0.096 | 0.097 |
| **Random Forest** | 0.348 | 0.393 | 0.348 | 0.338 |
| **SVM linear kernel** | 0.431 | 0.435 | 0.431 | 0.423 |
| **SVM RBF kernel** | 0.446 | 0.479 | 0.446 | 0.440 |

**Table 5** Summary of test results

# 6 CONCLUSION AND FUTURE WORK

In our work we have analyzed the performance of different recurrent models like RNN, LSTM and GRU after applying word embedding to the sequence data to classify the protein sequences to their respective families. We have also compared the results by applying trigram with deep neural network and shallow neural network. Neural networks are preferred over traditional machine learning models because they capture optimal feature representation by themselves taking the primary protein sequences as input and give considerably high family classification accuracy of about 96%.

Deep neural networks architecture is very complex therefore, understanding the background mechanics of a neural network model remain as a black box and thus the internal operation of the network is only partially demonstrated. In the future work, the internal working of the network can be explored by examining the Eigenvalues and Eigenvectors across several time steps obtained by transforming the state of the network to linearized dynamics[32].